# ENGINEERING MODELING OF WAVE-RELATED SUSPENDED SEDIMENT TRANSPORT OVER RIPPLES

RAFIK ABSI[1]

[1] *EBI, IPSL, Cergy University, 32 Boulevard du Port, 95094 Cergy-Pontoise France.* r.absi@ebi-edu.com

**Abstract**: The aim of our study is to improve the description of suspended sediment transport over wave ripples. We will first show the importance of sediment diffusivity with convective transfer (hereafter called $\varepsilon_s^*$) which is different from the sediment diffusivity $\varepsilon_s$ associated to turbulent flux $\overline{c'w'}$. It is possible to interpret concentration profiles, in semi-log plots, thanks to a relation between second derivative of the logarithm of concentration and derivative of $\varepsilon_s^*$ (Absi, 2010). An analytical profile for $\varepsilon_s^*$ will be presented and validated by experimental data obtained by Thorne *et al.* (2009) for medium sand. The proposed profile allows a good description of suspended sediment concentrations for fine and coarse sediments.

## Introduction

The description of sediment transport in the coastal zone is of crucial importance for accurate predictions of coast line evolution and sea-bed changes. The modeling of coastal sediment transport needs a compromise between two types of models: detailed mathematical models and engineering approaches. This compromise is imposed by on the one hand the accuracy of predictions and on the other hand the usability in practical applications. In coastal engineering, we need practical accurate engineering models which take into account the more important involved physics.

In the engineering approach, the net (averaged over the wave period) total sediment transport is obtained as the sum of the net bed load and net suspended load transport rates. For suspended load, the net sand transport is defined as the sum of the net current-related and the net wave-related transport components. The wave-related suspended sediment transport is defined as the transport of sediments by the high-frequency oscillatory flow or short waves (low-frequency transport is neglected).

The focus point of our study is the mixing of suspended sediments over wave ripples. The aim is to improve the prediction of wave-related suspended transport component in engineering modeling of coastal sediment transport.

The wave-related suspended transport component is modeled (Houwman and Ruessink, 1996; Grasmeijer *et al.*, 1999; van Rijn, 2003; van Rijn, 2007) as

$$q_{s,w} = \gamma \frac{U_{on}^4 - U_{off}^4}{U_{on}^3 + U_{off}^3} \int c \, dz \qquad (1)$$

where $U_{on}$ = the near-bed peak orbital velocity in onshore direction (in wave direction) and $U_{off}$ = the near-bed peak orbital velocity in offshore direction (against wave direction), c = the time-averaged concentration and γ = a phase lag parameter between 0.1 and 0.2. Eq. (1) requires computation of the time-averaged sand concentration profile and its integration in the vertical direction.

**Suspended Sediment Concentrations due to a Current**

Time-averaged suspended sediment concentrations, which result from the balance between an upward turbulent mixing flux and a downward gravity settling flux, are obtained from the advection-diffusion equation

$$w_s c + \varepsilon_s \frac{dc}{dz} = 0 \qquad (2)$$

where $w_s$ = sediments fall velocity (m/s), $\varepsilon_s$ = sediment mixing coefficient or sediment diffusivity (m²/s), and c = time-averaged concentration. Often suspended sediment concentration profiles are in semi-log plots upward concave and/or upward convex. In order to link this behaviour to sediment diffusivity, Absi (2010) wrote

$$\frac{d^2 \ln c}{d z^2} = \frac{w_s}{\varepsilon_s^2} \frac{d \varepsilon_s}{d z} \qquad (3)$$

Eq. (3) provides a link between $d^2 \ln c / d z^2$ and $d \varepsilon_s / d z$, and therefore between upward concavity/convexity of concentration profiles (in semi-log plots) and increasing/decreasing of sediment diffusivity profiles. Since $w_s / \varepsilon_s^2$ is always > 0; $d^2 \ln c / d z^2$ and $d \varepsilon_s / d z$ have the same sign and therefore increasing sediment diffusivity allows upward concave concentration profile, while decreasing sediment diffusivity allows an upward convex concentration profile (Absi, 2010).

*Turbulent mixing of suspended sediments*

In order to allow adequate predictions of suspended sediment transport, it is important to understand the effect of sediment particles on turbulence of fluid flow.

Rafik Absi                                                                           1097

The study of Gore & Crowe (1989) showed that small particles attenuate turbulence while large particles generate turbulence. Their analysis of experimental data suggested that when the ration between particle diameter and turbulence length scale is larger than about 0.1, particles change turbulence intensity of the fluid. Hetsroni (1989) suggested that the presence of particles with low particle Reynolds number (based on particle size and relative velocity) tends to suppress the turbulence while particles with high particle Reynolds number cause enhancement of turbulence. Elghobashi (1994) has classified fluid-solid two phase flows on the basis of the volumetric particle concentration $\alpha_p$ and Stokes number $St = \tau_p / \tau_t$ (where $\tau_p$ is the particle timescale and $\tau_t$ the integral turbulence timescale or turnover time of large eddy).

- For $\alpha_p < 10^{-6}$, the presence of particles have negligible effect on turbulence (one-way coupling)
- For $10^{-3} < \alpha_p < 10^{-6}$, the particles can augment the turbulence for Stokes number $St > 1$, or attenuate turbulence if $St < 1$ (two-way coupling).
- For $\alpha_p > 10^{-3}$, turbulence can be affected by particle-particle collisions (four-way coupling).

A large value of $St$ corresponds to a weak sedimentation process and therefore to a more uniform concentration profile. At the opposite, a small value of $St$ corresponds to a strong concentration gradient (Absi *et al.*, 2011). Rogers and Eaton (1990) defined the particle response time as

$$\tau_p = \frac{w_s}{\left(1 - \frac{\rho_f}{\rho_s}\right)g} \qquad (4)$$

where $\rho_f$ fluid density (kg/m$^3$), $\rho_s$ solid density (kg/m$^3$) and g the gravitational acceleration (m/s$^{-2}$). This timescale is the ratio of particles settling velocity to their vertical acceleration (vertical force – gravity minus buoyancy – per unit mass). The integral turbulence timescale (mixing time by the large eddies) is given by $\tau_t \approx \frac{l_m}{u_{rms}} \approx \frac{l_m}{\sqrt{k}}$, by $\tau_t \approx \frac{k}{\varepsilon}$ (where $k$ is the turbulent kinetic energy TKE, $\varepsilon$ is the dissipation of TKE), or by $\tau_t \approx \frac{\nu_t}{k}$ (Absi *et al.*, 2011).

Turbulent diffusion of suspended sediments $\varepsilon_s$ is modelled by $\varepsilon_s = \frac{1}{Sc_t} \nu_t$ or $\varepsilon_s = (1 + a\, Ri)^b \frac{1}{Sc_t} \nu_t$, where $\nu_t$ is the eddy viscosity, $Sc_t$ the turbulent Schmidt number, $Ri$ the Richardson number and *a* and *b* empirical coefficients.

Two formulations of turbulent Schmidt number $Sc_t$ based on a two-fluid description and a kinetic model were analysed by Absi *et al.* (2011). The kinetic model for turbulent two-phase flows provided $Sc_t$ which depends on particle Stokes number. The study showed that both approaches provided $Sc_t$ that depends on turbulent kinetic energy TKE, eddy viscosity and particles settling velocity.

In this study we write $\varepsilon_s$ as

$$\varepsilon_s = \beta \nu_t \qquad (5)$$

Or

$$\varepsilon_s = \beta \phi \nu_t^0 \qquad (6)$$

where $\beta$ = inverse of the turbulent Schmidt number, describes the difference between diffusivity of momentum (diffusion of a fluid "particle") and diffusivity of sediment particles, $\phi$ = parameter which accounts for the influence of sediments on the turbulence structure of the fluid (eddy viscosity), and $\nu_t^0$ is the eddy viscosity of fluid without sediments. For fine sediments, it is often assumed that $\varepsilon_s \approx \nu_t^0$ which is based on the hypothesis that fine sediments have no effect on eddy viscosity ($\phi = 1$) and the difference between diffusivity of momentum and diffusivity of fine particles is neglected ($\beta = 1$). When $\phi$ is assumed equal to 1, $\beta$ is the ratio of sediment diffusivity (with the presence of sand) to the momentum diffusivity in the absence of sand $\nu_t^0$.

The eddy viscosity is obtained by solving a two-equation turbulence model or by the following analytical formulation

$$\nu_t = A \kappa u_* z \exp(-C\, z/\delta) \qquad (7)$$

where $u_*$ = the friction velocity (m/s), $\delta$ = the boundary layer thickness (m), $\kappa$ = the Karman constant (=0.41) and A (=1) and C (=1.12) two parameters (Absi 2010).

The value of $\beta$ has been the subject of much research. Researchers found that $\beta$ approaches unity for fine sediments and deviates for coarse ones. In suspension flows over movable beds, experiments show that depth-averaged $\beta$-values are smaller than unity $\beta < 1$ for flows without bed forms while they are larger than unity $\beta > 1$ for flows with bed forms (Graf and Cellino, 2002). The finite-mixing-length model (Nielsen and Teakle, 2004) allows to obtain values of $\beta >$ or $<$ than 1 since

$$\beta = \frac{\varepsilon_s}{\nu_t} = \frac{1 + \frac{l_m^2}{24} L_c^{-2} + ...}{1 + \frac{l_m^2}{24} L_u^{-2} + ...} = 1 + \frac{l_m^2}{24}\left(L_c^{-2} - L_u^{-2}\right) + ... \qquad (8)$$

It is possible to obtain a theoretical $\beta(z)$ profile from Eq. (8) since $l_m$ is z-dependent. With a linear mixing length $l_m = \kappa z$, we wrote (Absi, 2010)

$$\beta = 1 + C_{\beta 2} z^2 \qquad (9)$$

where $C_{\beta 2} = \frac{\kappa^2}{24}\left(L_c^{-2} - L_u^{-2}\right) + ...$ . In order to allow analytical analysis, we suggested a simple equation (10) which fits Eq. (9)

$$\beta = \beta_b \exp\left(C_\beta z / \delta\right) \qquad (10)$$

where $\beta_b$ = the value of $\beta$ close to the bed and $C_b$ = coefficient. Using the $\beta$-function (10) and eddy viscosity (7), the dimensionless sediment diffusivity is given therefore by

$$\frac{\varepsilon_s}{U_0 k_s} = A_s \frac{z}{k_s} e^{-\frac{z/k_s}{B_s/k_s}} \qquad (11)$$

where $U_0$ = the maximum value of the free stream velocity (m/s) and $k_s$ the equivalent roughness (m), $A_s = A \kappa \beta_b \left(u_* / U_0\right)$ and $B_s = \delta / \left(C - C_\beta\right)$.

**Time-Averaged Concentrations over Wave Ripples**

In oscillatory flows, it is known that cycle-mean sediment diffusivity above ripples is significantly greater than the cycle-mean eddy viscosity, i.e. $\beta > 1$ (Nielsen, 1992; Thorne *et al.*, 2002). The value of $\beta$ was suggested empirically a constant equal to about four ($\beta = 4$) for rippled beds (Nielsen, 1992) and the near bed sediment diffusivity a constant equal to $\varepsilon_s = 0.016\, k_s U_0$. Another empirical sediment diffusivity formulation (van Rijn, 2007) involves a three-layer distribution (Fig. 1).

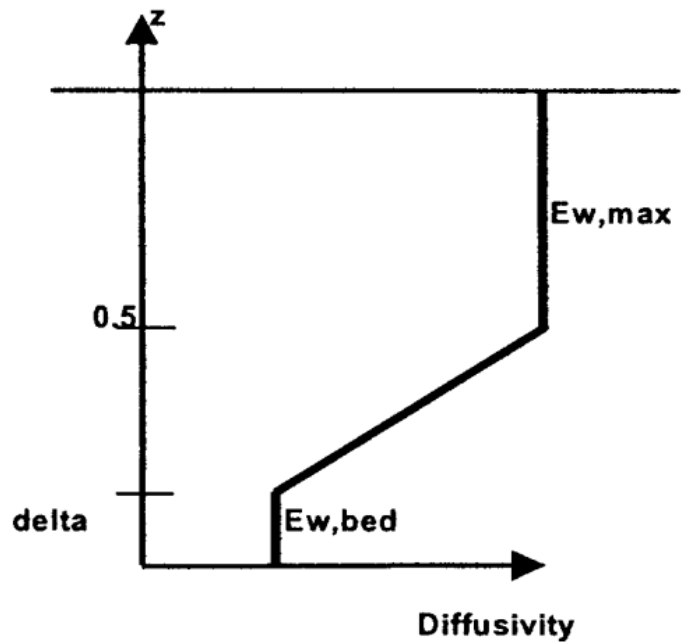

Fig. 1. Vertical distribution of wave-related sediment diffusivity (van Rijn, 2007)

Since sediment diffusivity $\varepsilon_s$ describes the disorganized ''diffusive'' process. The classical 1-DV gradient diffusion model (Eq. 2) seems unable to predict the process of vortex formation and shedding above ripples which is a coherent phenomenon. The process of vortex formation and shedding at flow reversal above ripples is a relatively coherent phenomenon. The associated convective sediment entrainment process may also be characterized as coherent, instead of a pure disorganized ''diffusive'' process represented in the classical gradient diffusion model (Thorne *et al.* 2002). Above ripples, in a ripple-averaged sense, the convective term can dominate the upward sediment flux in the bottom part of the wave boundary layer. Above this region the vortices lose their coherence and gradient diffusion becomes dominant characterized by a mixing length which grows and resulting in increasing sediment diffusivity with height above the bed (Thorne *et al.* 2009). Nielsen (1992) indicated that both convective and diffusive mechanisms are involved in the entrainment processes. In the combined convection-diffusion formulation, the steady state advection-diffusion equation is given by

$$w_s c + \varepsilon_s \frac{dc}{dz} + F_{conv} = 0 \tag{12}$$

The respective terms in (12) represent downward settling, upward diffusion (given by gradient diffusion $F_{diff} = \varepsilon_s (dc/dz)$) and upward convection $F_{conv}$. The upward convection term $F_{conv}$ was given by Nielsen (1992) as $F_{conv} = -w_s c_0 F(z)$, where F(z) is a function describing the probability of a particle reaching height z above the bed (Nielsen, 1992; Thorne *et al.*, 2002; Lee and Hanes, 1996). Thorne *et al.* (2009) wrote $F_{conv} = -\overline{w_w c_w}$ where $w_w$ and $c_w$ are periodic components respectively of concentrations and vertical velocity and the overbar denotes time averaging. It is possible to write (12) in the form of diffusion equation (Eq. 2). The time-averaged (over the wave period) advection-diffusion equation is given therefore by (Absi, 2010)

$$w_s c + \varepsilon_s^* \frac{dc}{dz} = 0 \tag{13}$$

where $\varepsilon_s^* = \alpha \varepsilon_s$ and $\alpha$ = a parameter related to convective sediment entrainment process associated to the process of vortex shedding above ripples $\alpha = 1/(1 + F_{conv}/(w_s c))$. With the upward convection $F_{conv} = -w_s c_0 F(z)$ (Nielsen, 1992), $\alpha$ becomes equal to $1/(1 - (c_0/c)F(z))$, while with $F_{conv} = -\overline{w_w c_w}$ (Thorne *et al.*, 2009) $\alpha = 1/(1 - \overline{w_w c_w}/(w_s c))$. The condition of Sheng and Hay (1995) $\overline{w_w c_w}/(w_s c) < 0.2$ shows therefore that when the convective transfer is very small (above low steepness ripples), $\alpha \approx 1$ and therefore $\varepsilon_s^* \approx \varepsilon_s$. From equations (12) and (13), it is possible to write

$$\varepsilon_s^* = \left(1 + \frac{F_{conv}}{\varepsilon_s \frac{dc}{dz}}\right) \varepsilon_s \tag{14}$$

and therefore $\alpha = 1 + (F_{conv}/(\varepsilon_s (dc/dz))) = 1 + (F_{conv}/F_{diff})$. This equation shows that $\alpha$ depends on the relative importance of coherent vortex shedding (related to $F_{conv}$) and random turbulence (related to $F_{diff}$). When $F_{conv} > F_{diff}$ => $\alpha > 1$, while $F_{conv} \ll F_{diff}$ => $\alpha \approx 1$ and therefore $\varepsilon_s^* \approx \varepsilon_s$. With Eq. (13), Eq. (3) becomes

$$\frac{d^2 \ln c}{dz^2} = \frac{w_s}{\varepsilon_s^{*2}} \frac{d\varepsilon_s^*}{dz} \tag{15}$$

Eq. (15) provides a link between upward concavity/convexity of concentration profiles (in semi-log plots) and increasing/decreasing of $\varepsilon_s^*$. Increasing $\varepsilon_s^*$ allows upward concave concentration profile, while decreasing $\varepsilon_s^*$ allows an upward convex concentration profile. Absi (2010) suggested an empirical function for $\alpha$ given by

$$\alpha = 1 + D \exp\left(-z/h_s\right) \tag{16}$$

where D and $h_s$ are two parameters.

**Results and discussion**

Two test cases are presented. The first concerns fine and coarse sediments over wave ripples in the same flow while the second case presents acoustic measurements of near-bed sediment diffusivity profiles over two sandy rippled beds (medium and fine in term of sand grain size) under waves.

*Fine and coarse sediments over wave ripples in the same flow (McFetridge and Nielsen, 1985)*

Figure (2) shows fine and coarse sediments over wave ripples in the same flow. These experimental data show a particular interest: Since fine and coarse sediments are in the same flow, turbulence structure of the fluid is influenced by both sediments and the parameter $\phi$ is therefore the same. The difference in sediment diffusivity between fine and coarse sediments is related therefore only to parameter $\beta$. Eqs. (11) and (16) allows to write $\varepsilon_s^*$ as

$$\frac{\varepsilon_s^*}{U_0 k_s} = A_s \frac{z}{k_s} e^{-\frac{z/k_s}{B_s/k_s}} \left(1 + D\, e^{-\frac{z/k_s}{h_s/k_s}}\right) \tag{17}$$

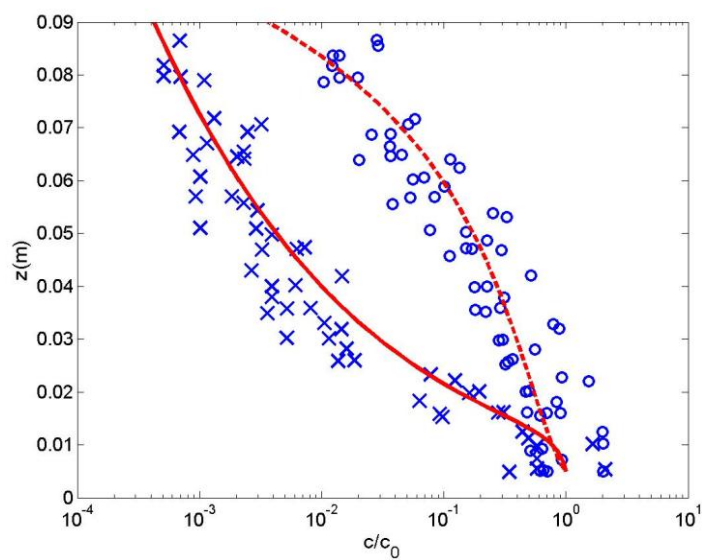

Fig. 2. Time-averaged concentration profiles over wave ripples. Symbols, measurements (McFetridge and Nielsen, 1985), (o) fine sand ($w_s$=0.65 cm/s); (x) coarse sand ($w_s$=6.1 cm/s). Curves are solutions of Eq. (12) with Eq. (16) (values of parameters are given in Absi 2010).

Experimental data (symbols) show a contrast between an upward convex concentration profiles, time-averaged in semi-log plots, for fine sand (o) and an upward concave profiles for coarse sand (x). Careful examination of data for coarse sand shows a near-bed upward convex profile beneath the main upward concave profile. The 1-DV gradient diffusion model (Eq. 2) with a constant settling velocity ($w_s$ = cste) and $\beta = 1$, predicts concentration profile for fine sediments (dashed line in fig. 2) but fails for coarse sand. The 1-DV gradient diffusion model predicts the main upward concave profile for coarse sediments thanks to β-function of Eq. (10) which was validated by the finite-mixing-length model. The model with the resulting sediment diffusivity $\varepsilon_s$ and a constant settling velocity is unable to predict the near-bed upward convex profile. In order to predict this profile, an additional parameter α is needed. This parameter was first related to settling velocity (α equal to inverse of dimensionless settling velocity). However, the dimensionless settling velocity decreases from z = 4 cm and at z = 2 cm the decreasing in settling velocity is of 50%. This is outside the range of observed hindered settling. Deeper analysis shows that this parameter should be related to convective sediment entrainment process and provides $\varepsilon_s^*$. Equations (13) with (17) allow very good description (solid line in fig. 2) of coarse sediments.

These profiles are interpreted by a relation between second derivative of the logarithm of concentration and derivative of the product between sediment diffusivity and α (Eq. 15).

*Medium and fine sediments over wave ripples (Thorne et al. 2009)*

Thorne *et al*. (2009) presented an analysis of observed near-bed sediment diffusivity profiles over sandy rippled beds under waves. Sediment diffusivities were inferred from concentration profiles measured over two sandy rippled beds comprising medium and fine sand, under slightly asymmetric regular waves. In the medium sand case, the ripples had relatively steep slopes. The form of the sediment diffusivity profiles was found to be constant with height above the bed to a height equal approximately to the equivalent roughness of the bed $k_s$. Above this level the sediment diffusivity $\varepsilon_s$ increased linearly with height. For the case of the fine sand, the ripples slopes were approximately half that of the medium sand. There was no constant region in the sediment diffusivity profiles; the sediment diffusivity simply increased linearly with height from the bed. In the case of the medium sand, the constant value of sediment diffusivity close to the bed was related to coherent vortex shedding. Steep ripples involve flow separation on the lee side of ripple crest and vortex formation. The vortices lose their coherence above the vortex layer (where $\varepsilon_s^* =$ constant) and gradient diffusion becomes dominant, characterized by increasing sediment diffusivity. For the fine sand case and low slope ripples, the bed is considered as dynamically plane. No near-bed $\varepsilon_s$ constant layer was observed because no flow separation or vortex formation occurs. Random turbulent processes explain the observed linear form for $\varepsilon_s$.

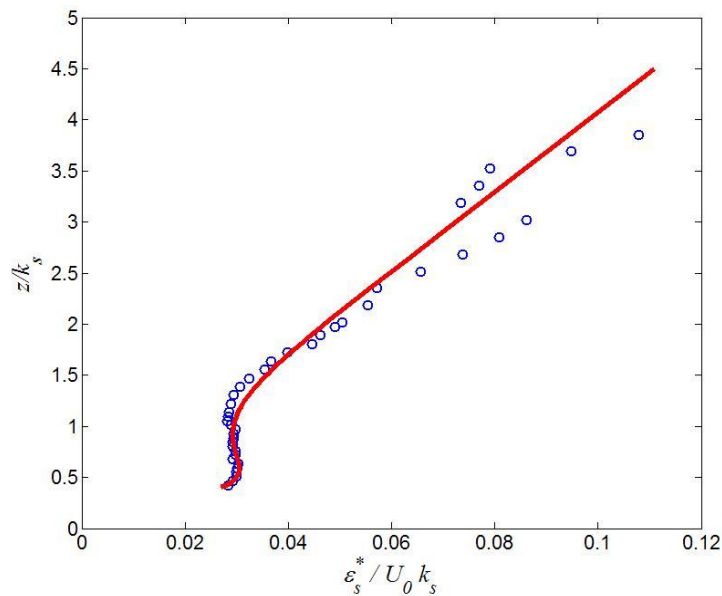

Fig. 3. Comparison of the mean measured normalized $\varepsilon_s^*$ (symbols) over medium sand bed (Thorne *et al*. 2009), with calculations from equation (18) (curve) with parameters $A_s$=0.026, $z_0/k_s$=0.2, $B_s/k_s$=500, D=9 and $h_s/k_s$=0.26.

Rafik Absi     1105

Figure (3) shows mean measured normalized $\varepsilon_s^*$ (symbols) over medium sand bed and comparison with our analytical profile. In Eq. (16), z is the height above ripple crest. However, concentration profiles of Thorne *et al.* (2009) were referenced to the undisturbed bed. Since z is the height above the undisturbed bed, we write Eq. (17) as

$$\frac{\varepsilon_s^*}{U_0 k_s} = A_s \frac{(z-z_0)}{k_s} e^{-\frac{(z-z_0)/k_s}{B_s/k_s}} \left(1 + D\, e^{-\frac{(z-z_0)/k_s}{h_s/k_s}}\right) \quad (18)$$

where $z_0$ is the distance between the undisturbed bed level and ripple crest. Equation (18) contains two different contributions given respectively by parameters $\beta$ (i.e. the inverse of the turbulent Schmidt number) and $\alpha$ (related to convective transfer). Comparison of the mean measured normalized $\varepsilon_s^*$ obtained by Thorne *et al.* (2009) (symbols) over the medium sand bed, with calculation (curve) from equation (18) shows good agreement. The shape is similar to the lower part of the profile of van Rijn (2007) (figure 1).

**Conclusions**

The findings of the present study can be summarized in the following conclusions:

- Adequate predictions of suspended sediment transport over wave ripples depend on sediment diffusivity with convective transfer $\varepsilon_s^*$ which is different from the sediment diffusivity $\varepsilon_s$ associated to turbulent flux $\overline{c'w'}$

- It is possible to interpret concentration profiles, in semi-log plots, thanks to a relation between second derivative of the logarithm of concentration and derivative of $\varepsilon_s^*$

- We presented a profile for $\varepsilon_s^*$, based on a β-function validated by the finite-mixing-length model and an additional parameter α related to convective sediment entrainment process associated to the process of vortex shedding above ripples

- The proposed profile for $\varepsilon_s^*$ allows a good description of suspended sediment concentrations for fine and coarse sediments (in the same flow) over wave ripples. The 1-DV gradient diffusion model predicts respectively the main upward concave profile for coarse sediments thanks to the β-function and the near-bed upward convex profile thanks to α

Rafik Absi 1106

- Comparison between the proposed profile for $\varepsilon_s^*$ and experimental data of Thorne *et al*. (2009) shows good agreement

- The profile for $\varepsilon_s^*$ needs more analysis and a calibration for predictive purpose

**Acknowledgements**

The author would like to thank Prof. Peter D. Thorne and Prof. Peter Nielsen for having provided the experimental data files. He would like also to thank Prof. Daniel C. Conley for his useful comments.

**References**

Absi, R. (2010). "Concentration profiles for fine and coarse sediments suspended by waves over ripples: An analytical study with the 1-DV gradient diffusion model," *Advances in Water Resources,* Elsevier, 33(4): 411–418.

Absi, R., Marchandon, S., and Lavarde, M. (2011), "Turbulent diffusion of suspended particles: analysis of the turbulent Schmidt number," *Defect and Diffusion Forum*, Trans Tech Publications, in press.

Elghobashi, S. E. (1994). "On predicting particleladen turbulent flows," *Appl. Sci. Res.*, 52, 309-329.

Gore, R.A., and Crowe, C.T. (1989). "Effect of particle size on modulating turbulence intensity," *Int. J. Multiphase Flow*, 15, 279-285.

Graf, W.H., and Cellino, M. (2002). "Suspension flows in open channels: experimental study," *J. Hydraul. Res.*, 40(4), 435–47.

Grasmeijer, B. T., Chung, D. H., and van Rijn, L. C. (1999). "Depthintegrated sand transport in the surf zone," *Proc., Coastal Sediments*, ASCE, Reston, Va., 325–340.

Hetsroni, G. (1989). "Particles-turbulence interaction," *Int. J. Mulriphase Flow*, 15, 735-746.

Houwman, K. T. and Ruessink, B. G. (1996). "Sediment transport in the vicinity of the shoreface nourishment of Terschelling," *Report,* Dep. of Physical Geography. University of Utrecht, The Netherlands.

Lee, T.H., and Hanes, D.M. (1996). "Comparison of field observations of the vertical distribution of suspended sand and its prediction by models," *J. Geophys. Res.*, 101, 3561-3572.

McFetridge, W. F., Nielsen, P. (1985). "Sediment suspension by non-breaking waves over rippled beds," *Technical Report* No. UFL/COEL-85/005, Coast Ocean Eng Dept, University of Florida.

Nielsen, P. (1992). "*Coastal bottom boundary layers and sediment transport,*" World Scientific, 324 p.

Nielsen, P, and Teakle, I.A.L. (2004). "Turbulent diffusion of momentum and suspended particles: a finite-mixing-length-theory," *Phys. Fluids*, 16(7), 2342–8.

Rogers, C.B., and Eaton, J.K. (1990). "The behavior of solid particles in a vertical turbulent boundary layer in air," *International Journal of Multiphase Flow*, 16(5), 819-834

Sheng, J., and Hay, A. E. (1995). "Sediment eddy diffusivities in the nearshore zone, from multifrequency acoustic backscatter", *Cont. Shelf Res.*, 15(2–3), 129-147, doi:10.1016/0278-4343(94)E0025-H.

Thorne, P.D., Williams, J.J., and Davies, A.G. (2002). "Suspended sediments under waves measured in a large-scale flume facility," J. Geophys. Res., 107(C8), 3178.

Thorne, P. D., Davies, A. G., Bell, P. S. (2009). "Observations and analysis of sediment diffusivity profiles over sandy rippled beds under waves," *J. Geophys. Res.*, 114:C02023.

van Rijn, L. C. (2003). "Sand transport by currents and waves: general approximation formulae," *Proceedings Coastal Sediments '03*, Clearwater Beach, Florida, USA, CD-ROM Published by East Meets West Productions, Corpus Christi, TX ISBN-981-238-422-7, 14 p.

van Rijn, L.C. (2007). "United view of sediment transport by currents and waves II: Suspended transport," *Journal of Hydraulic Engineering*, ASCE, Vol. 133, No. 6, p. 668-689.